\documentclass[useAMS,usenatbib]{mn2e}
\usepackage{epsfig}
\usepackage{threeparttable}
\usepackage{amsmath}
\title[Constraining the dipolar magnetic field of M82 X-2]{Constraining the dipolar magnetic field of M82 X-2 by the accretion model}
\author[W. C. Chen]{ Wen-Cong Chen $^{1,2}$\thanks{E-mail:
chenwc@pku.edu.cn}\\
$^1$ School of Physics and Electrical Information, Shangqiu Normal University,
Shangqiu 476000, China; \\
$^2$ Department of Physics, University of Oxford, Oxford OX1 3RH, UK;\\
 }
\begin{document}

\date{}

\pagerange{\pageref{firstpage}--\pageref{lastpage}} \pubyear{2011}

\maketitle

\label{firstpage}

\begin{abstract}
Recently, ultraluminous X-ray source (ULX) M82 X-2 has been identified to be an accreting neutron star, which has a $P=1.37$ s spin period, and is spinning up at a rate $\dot{P}=-2.0\times 10^{-10}~\rm s\,s^{-1}$. Interestingly, its isotropic X-ray luminosity
$L_{\rm iso}=1.8\times 10^{40}~\rm erg\,s^{-1}$ during outbursts is 100 times the Eddington limit for a $1.4~\rm M_{\odot}$ neutron star. In this Letter, based on the standard accretion model we attempt to constrain the dipolar magnetic field of the pulsar in ULX M82 X-2. Our calculations indicate that the accretion rate at the magnetospheric radius must be super-Eddington during outbursts. To support such a super-Eddington accretion, a relatively high multipole field ($\ga 10^{13}$ G) near the surface of the accretor is invoked to produce an accreting gas column. However, our constraint shows that the surface dipolar magnetic field of the pulsar should be in the range of $1.0-3.5\times 10^{12}$ G. Therefore, our model supports that the neutron star in ULX M82 X-2 could be a low magnetic field magnetar (proposed by Tong) with a normal dipolar field ($\sim 10^{12}$ G) and relatively strong multipole field.  For the large luminosity variations of this source, our scenario can also present a self-consistency interpretation.
\end{abstract}

\begin{keywords}
X-rays: binaries -- stars: magnetic field -- stars:magnetar -- X-rays:bursts -- stars: neutron
-- X-rays: individual (NuSTAR J095551+6940.8)
\end{keywords}

\section{Introduction}
Ultraluminous X-ray sources (ULXs) are defined as point extranuclear sources found in nearby galaxies. ULXs are characterized by
isotropic X-ray luminosities exceeding $\sim10^{40}~\rm erg\,s^{-1}$ (some papers adopted a luminosity of $10^{39}~\rm erg\,s^{-1}$), which is larger than the Eddington limit of a $10~\rm M_{\odot}$ stellar mass black hole \citep{fabb89}. To interpret the extreme luminosities, some works suggested that ULXs originated from stellar mass black holes by either actually super-Eddington X-ray radiation \citep{bege02} or anisotropic beaming emission \citep{king01,kord02}. Alternatively, the accretors in some ULXs are proposed to be intermediate-mass black holes ($10^{2}-10^{4}~ \rm M_{\odot}$) \citep{colb99,li04}.

Since higher masses naturally yield a larger isotropic Eddington limit, theoretical models have favored black hole rather than neutron star X-ray binaries. The dynamical measurements for two ULXs masses support this viewpoint \citep{liu13,motc14}. Recently, the identification of NuSTAR J095551+6940.8 in the nearby galaxy M82 has raised an extreme interest of theoretical researchers \citep{bach14}. The broadband X-ray observations for M82 X-2 (also entitled as X42.3+59, discovered by Kaaret, Simet \& Lang 2006) imply pulsations with an average period of 1.37 s and a 2.5 d sinusoidal modulation. The former comes from the spin of a magnetized neutron star, and the latter is explained as the orbital motion. The isotropic X-ray luminosity in the 0.3 - 10 keV can reach
$L_{\rm iso}=1.8\times 10^{40}~\rm erg\,s^{-1}$, which is 100 times the Eddington limit for a $\rm 1.4~M_{\odot}$ neutron star. Meanwhile, the pulsar in M82 X-2 was reported to be spinning up at a rate $\dot{P}=-2.0\times 10^{-10}~\rm s\,s^{-1}$ (in the interval from MJD 56696 to 56701) in a high-mass X-ray binary including a donor star with mass greater than $5.2~\rm M_{\odot}$ \citep{bach14}.

Simulations by binary population synthesis show that the contribution of neutron star X-ray binaries for the ULXs population obviously exceeds black hole X-ray binaries \citep{shao15}. It is an intriguing issue how to radiate such an X-ray luminosity for an accreting neutron star. \cite{eksi15} show that the pulsar in M82 X-2 is a magnetar with a dipole magnetic field of $6.7\times 10^{13}$ G, which can reduce the scattering cross-section and enhance the critical luminosity \citep{pacz92}. Based on the observational properties of M82 X-2, the calculation for 30 keV photons performed by \cite{dall15} indicates that a magnetic field $\sim 10^{13}$ G can reduce the scattering cross-section by a factor of 50. \cite{kari16} have simulated the formation of M82 X-2, and concluded that a neutron star with $4\times10^{13}$ G magnetic field accreting from a disc-shape wind produced by a Be-Companion can reproduce the observed parameters. Conversely, \cite{kluz15} favored a neutron star with a weak magnetic field of $10^{9}$ G. They argued that such a magnetic field allows the accretion disc to extend to the surface of the neutron star, providing the spin-up torque fitting the observed properties. \cite{tong15} argued that the accretion column originating from strong multipole field of a low magnetic field magnetar could responsible for the super-Eddington accretion. Recently, \cite{king16} presented an interesting picture, in which M82 X-2 is a beamed X-ray source feeding at super-Eddington rate.

In present, the surface dipolar magnetic field and large X-ray luminosity variation in M82 X-2 still remain controversial.
Based on the standard accretion model, in this Letter we attempt to constrain the surface dipole magnetic field and the beaming factor of M82 X-2.

\section{Accretion model}
We assume that M82 X-2 is a high-mass X-ray binary where the magnetic neutron star is accreting material from a donor star overflowing Roche lobe via an accretion disc. In this case, the magnetic accretion process strongly depends on the magnetospheric
(Alfv\'{e}n) radius $r_{\rm m}$, at which the ram pressure of the accreting material and the magnetic pressure are equal. The magnetospheric radius is given by \citep{davi73}
\begin{equation}
r_{\rm m}=\xi\left(\frac{\mu^{4}}{2GM\dot{M}^{2}}\right)^{1/7},
\end{equation}
where $G$ is the gravitational constant, $\xi$ a dimensionless parameter of the order unity, $\dot{M}$ the accretion rate at $r_{\rm m}$, $M$ and $\mu=B_{\rm p}R^{3}$ ($B_{\rm p}$ and $R$ are the surface dipole magnetic field and the radius of the neutron star) the mass and the dipolar magnetic momentum of the neutron star, respectively. Inserting some typical parameters in equation (1), thus
\begin{equation}
r_{\rm m}=1.6\times 10^{8}\xi \dot{M}^{-2/7}_{18}M^{-1/7}_{1.4}\mu^{4/7}_{30}~\rm cm,
\end{equation}
where $\dot{M}_{18}$ in units of $10^{18}~\rm g\,s^{-1}$, $M_{1.4}$ in units of $1.4~\rm M_{\odot}$, and $\mu_{30}$ in units of $10^{30}~\rm G\,cm^{3}$.

Mass and angular momentum accretion can spin up the neutron star. Meanwhile, there also exists a spin-up torque originating from the magnetic coupling between the neutron star and the inner region the accretion disc. Similarly, there is a spin-down torque acting on the neutron star by the outer disc \citep{ghos79b,rude89}. However, both magnetic torques depend on the so-called fastness parameter $\omega_{\rm s}=\Omega/\Omega(r_{\rm m})$ (where $\Omega$, and $\Omega(r_{\rm m})$ are the angular velocity of the neutron star, and the Keplerian angular velocity at $r_{\rm m}$, respectively). Because it is difficult to calculate the fastness parameter without the magnetic field, we ignore the contribution of the magnetic torque \footnote{Equation (10) of \cite{ghos79b} shows that the total spin-up torque is at most 1.39 times the accretion torque. }. Therefore, we approximately have
\begin{equation}
-\frac{2\pi I\dot{P}}{P^{2}}\leq \dot{M}\sqrt{GMr_{\rm m}},
\end{equation}
where $I$ is the moment of inertia of the neutron star.
Using the observed parameters $P=1.37$ s, $\dot{P}=-2\times 10^{-10}~\rm s\,s^{-1}$ \citep{bach14}, and taking $I=10^{45}~\rm g\,cm^{-2}$, it yields the minimum dipole magnetic momentum
\begin{equation}
\mu_{\rm min,30}=114\xi^{-7/4}\dot{M}^{-3}_{18}M^{-3/2}_{1.4}~\rm G\,cm^{3}.
\end{equation}

According to the accretion theory, the magnetospheric radius in the accretion phase should be smaller than the corotation radius
\begin{equation}
r_{\rm co} =\sqrt[3]{\frac{GMP^{2}}{4\pi^{2}}}=2.1\times 10^{8}M_{1.4}^{1/3}~\rm cm.
\end{equation}
Assuming $r_{\rm m}=r_{\rm co}$, one can obtain the maximum dipole magnetic momentum as follows
\begin{equation}
\mu_{\rm max,30}=1.6\xi^{-7/4}\dot{M}^{1/2}_{18}M^{5/6}_{1.4}~\rm G\,cm^{3}.
\end{equation}

Considering $\mu_{\rm max,30}\geq\mu_{\rm min,30}$, and taking $M_{1.4}=1.0$, the minimum accretion rate at the
magnetospheric radius must satisfy
\begin{equation}
\dot{M}_{\rm min}= 3.4\times 10^{18}~{\rm g\,s^{-1}}\simeq 3.4~\dot{M}_{\rm Edd},
\end{equation}
where $\dot{M}_{\rm Edd}=1.0\times 10^{18}M_{1.4}~\rm g\,s^{-1}$ is the Eddington accretion rate for steady spherical accretion \citep{fran02}. It is worth noting that this minimum accretion rate is independent on $\xi$. Therefore, the neutron star in M82 X-2 should experience a super-Eddington accretion.

For a strong magnetic neutron star, the material accreting along the magnetic field lines would produce an accretion column near the polar cap \citep{shap83}. Because the accreting material were frozen in the close magnetic lines between $r_{\rm m}$ and the light cylinder radius, it would form an accreting column (like a thin accreting 'wall') on the surface of the neutron star, in which the cross-section of the accretion column is like a narrow circular ring (see also Figure 1a of Basko \& Sunyaev 1976). The accretion energy were radiated from the side wall, so the accretion rate can exceed the Eddington rate \citep{bask76}.
In the accretion column case, the maximum accretion of a neutron star can be written as \citep{bask76}
\begin{equation}
\dot{M}_{\rm max}= \frac{l_{0}}{2\pi d_{0}}\dot{M}_{\rm Edd}\simeq 6.4\times 10^{18}(\frac{l_{0}/d_{0}}{40})M_{1.4}~\rm g\,s^{-1} ,
\end{equation}
where $l_{0}$, and $d_{0}$ are the length, and the thickness of the accretion column, respectively. The radius of the narrow circular ring on the surface of the neutron star can be written as
\begin{equation}
r_{\rm column}= R\sqrt{\frac{R}{r_{\rm m}}}.
\end{equation}
Due to $r_{\rm column}\propto B^{-2/7}$, and $r_{\rm column}\propto \dot{M}^{1/7}$, it is not sensitive with the magnetic field and accretion rate. Therefore, taking $\dot{M}_{18}=3.4,~ M_{1.4}=1.0,~ R=10^{6}~\rm cm, ~\xi=1$, and $\mu_{30}=10$, we have $r_{\rm column}=4.9\times10^{4}~\rm cm$, and $l_{0}=2\pi r_{\rm column}=3.1\times10^{5}~\rm cm$. Considering a thin accreting 'wall' (its thickness $d_{0}\ll r_{\rm column}$) $d_{0}=10^{4}~\rm cm$ (see also Mushtukov et al. 2015 ), so we have $l_{0}/d_{0}\approx30$. Therefore, we can obtain the maximum accretion $\dot{M}_{\rm max}=4.8\times 10^{18}~\rm g\,s^{-1}$ from equation (8).

If the neutron star is accreting material at a rate $\dot{M}= 4.8\times 10^{18}~\rm g\,s^{-1}$, taking $\xi=1.0$ and the radius of the neutron star $R=10^{6}~\rm cm$, from the equations (6) and (4), the maximum and minimum of the surface dipole magnetic field can be constrained to be
\begin{equation}
B_{\rm p,max}=3.5\times 10^{12}~{\rm G}, ~B_{\rm p,min}=1.0\times 10^{12}~{\rm G}.
\end{equation}
If we take a $\xi=0.5$, two magnetic fields would increase by a factor of 3.4 (see also the bold curves in Figure 1).

\begin{figure}
 \includegraphics[width=0.55\textwidth]{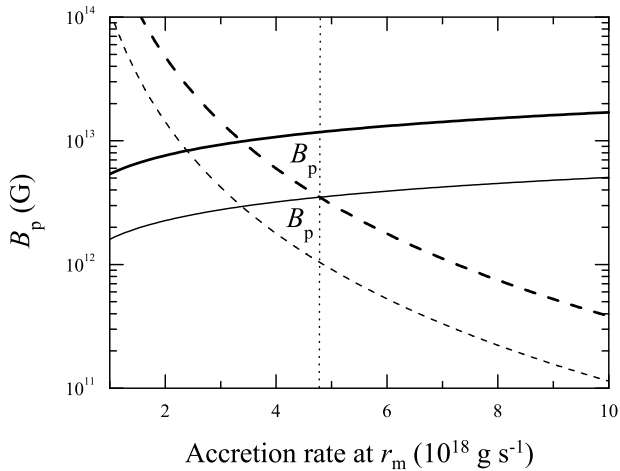}
  \caption{Relation between the surface dipole magnetic field of the neutron star and the accretion rate at the
magnetospheric radius. The solid curve, and dashed curve represent the maximum, and the minimum of the surface magnetic field of the neutron star (the thin, and bold curves correspond to $\xi=1.0$, and 0.5, respectively), respectively. The vertical dotted line denotes the upper limit of the accretion rate $\dot{M}_{\rm max}=  4.8\times 10^{18}~\rm g\,s^{-1} $.}
 \label{fig:fits}
\end{figure}

In Figure 1, we illustrate the relation between the surface dipole magnetic field of the neutron star and the accretion rate at the magnetospheric radius. The area enclosing by the solid curve, the dashed curve, and the dotted line represents a reasonable region of the dipole magnetic field of M82 X-2. It seems that the dipole magnetic field of the neutron star locates a normal range when $\xi=1.0$ (the discussion when $\xi=0.5$ see also section 3). However, to support such an accreting gas column, the accretor should have a relatively high multipole field ($\sim2\times10^{13}$ G) near the surface \citep{bask76}. Therefore, the neutron star in M82 X-2 could be a low magnetic field magnetar \citep{tong15}, which has a normal dipole magnetic field, and a strong surface multipole field \citep{turo11}.

In the low magnetic field magnetar framework, the maximum, and the minimum accretion rate are $4.8\times 10^{18}$, and $3.4\times 10^{18}~\rm g\,s^{-1}$, respectively; corresponding to the true maximum, and minimum X-ray luminosity $L_{\rm X, max}=8.6\times 10^{38}~\rm erg\,s^{-1}$, and $L_{\rm X, min}=6.1\times 10^{38}~\rm erg\,s^{-1}$. To account for the apparent isotropic X-ray luminosity $L_{\rm iso}=1.8\times 10^{40}~\rm erg\,s^{-1}$ (0.3 - 10 keV) of M82 X-2 in high state \citep{bach14}, a beaming radiation of the neutron star is expected. From $L_{\rm iso}=L_{\rm X}/b$ (where $b$ is the beaming factor), we can obtain the beaming factor range at peak luminosity as follows
\begin{equation}
0.03\leq b \leq 0.05.
\end{equation}
This constraint is slightly lower than the observational estimation ($b=0.2$) given by \cite{feng11}.

\section{Discussion and Conclusion}

Based on the observed properties of M82 X-2, using the standard accretion model we found the pulsar should experienced a super-Eddington accretion at a rate of at least $3.4~\dot{M}_{\rm Edd}$ during outbursts. Strong multipole field ($\ga ~10^{13}$ G) near the surface of the pulsar can result in an accretion gas column, which allow the accretor accreted the material at a super-Eddington rate. However, our calculation show that the upper limit of the dipole magnetic field in the pulsar pole is $3.5\times10^{12}$ G {when $\xi=1.0$. Therefore, we proposed that the pulsar in M82 X-2 could be low magnetic field magnetar. It has a normal dipole field ($\sim 10^{12}$ G) and relatively strong multipole field in the surface of the pulsar \citep{tong15}. At present, three candidates for low magnetic field magnetar including SGR 0418+5729 \citep{rea10}, Swift J1822.3$-$1606 \citep{rea12}, 3XMM J185246.6+003317 \citep{zhou14,rea14} have been detected. Based on the accretion induced the polar magnetic field decay model, the simulation given by \cite{pan16} show that the polar magnetic field decays to $4.5\times10^{13}$ G when the pulsar in M82 X-2 accreted $\sim 0.005~\rm M_{\odot}$, while the strong magnetic field still remain in the out-polar region.

According to the observed parameters of M82 X-2, and adopting a typical accretion column factor $l_{0}/d_{0}=30$, the dipole magnetic field in the pulsar pole and the beaming factor at peak luminosity are constrained to be $B_{\rm p}=1.0-3.5\times 10^{12}$ G, and $b= 0.03 - 0.05$, respectively. A high accretion rate would obtain a small minimum magnetic field.

Recently, based on the observations in the lowest observed states \citep{brig16,tsyg16}, \cite{chri16} used two independent methods to calculate the magnetic field of the pulsar in M82 X-2, and obtain two similar results of $3.1\times10^{12}$, and $2.3\times10^{12}$ G. Following their assumption, if the pulsar is in the propeller line, and is accreting at Eddington accretion rate, our equation (6) derives a magnetic field of $1.6\times10^{12}$ G (The main difference originates from different magnetic momentum ($\mu$) formulas: we use $\mu=BR^{3}$, while they took $\mu=BR^{3}/2$). However, in our scenario such an accretion rate and a magnetic field can not provide an accretion torque fitting the observation.

Due to the uncertainty of $\xi$ factor in the magnetospheric radius equation, it seems that our scenario can not fully rule out the magnetar possibility for M82 X-2. If we take a relatively small $\xi=0.5$ favored by \cite{ghos79a}, it would give rise to an upper limit of $1.2\times10^{13}$ G, which is still less than the quantum critical limit $B_{\rm c}=m_{\rm e}^{2}c^{3}/\hbar e=4.4\times 10^{13}$ G. However, our estimation for the accretion rate strongly depend on the accretion column model. If the pulsar in M82 X-2 is experiencing a hyper-critical accretion at a rate of 100 $\dot{M}_{\rm Edd}$, equation (6) shows that the maximum dipole magnetic field may reach $10^{14}$ G. It is worth noting that it just is an upper limit of the dipole magnetic field. In our opinion, at present it is difficult to make an accurate constraint for the dipole field of the pulsar in M82 X-2.

 \cite{kong07} detected M82 X-2 turned off twice in 1999, and 2000. Recently, \cite{dall16} reported that this source has not been detected in the first observation performed by \emph{Chandra} HRC. An analysis for \emph{Chandra} fifteen years data indicated that M82 X-2 has a violent X-ray luminosity variations from $10^{38}$ to $10^{40}~\rm erg\,s^{-1}$ \citep{brig16}. These observations indicate that M82 X-2 experienced a large variation in the accretion rate. In our model, only if the accretion rate at $r_{\rm m}$ during outbursts is greater than $3.4~\dot{M}_{\rm Edd}$ (see also equation 7), $r_{\rm m}<r_{\rm co}$, and accretion occurs.  Actually, the spin-up rate of M82 X-2 in the interval from MJD 56685.5 to 56692 is $\dot{P}\approx-4.0\times 10^{-11}~\rm s\,s^{-1}$ (see also Figure 2 of Bachetti et al. 2014), which also reveals that the accretion rate of the pulsar decreases by a factor of five. Therefore, the accretion rate should be greater than $\sim0.7~\dot{M}_{\rm Edd}$. When the accretion rate drops to less than $0.7~\dot{M}_{\rm Edd}$, $r_{\rm m}>r_{\rm co}$, the pulsar is in the propeller phase \citep{illa75}, and the accretion ceases or sub-Eddington accretion occurs. The disc inner radius in M82 X-2 has been reported to be $3.5^{+3.0}_{-1.9}\times 10^{9}$ cm \citep[90\% confidence,][]{feng10}.
This radius obviously exceed  the corotation radius, the standard accretion theory suggested that the accretion should halt because of centrifugal force. If this measurement is confident, meantime the pulsar should be in propeller phase.Some other neutron star X-ray binaries emitted much lower X-ray luminosities, which provided strong evidence of propeller phase \citep{cui97,zhan98}. \cite{dall16} proposed a similar model, while their critical mass inflow rate is $\dot{M}_{\rm Edd}$. The sub-Eddington accretion would cause the source luminosity to decline by a large factor. If the accretion fully stops, much smaller X-ray luminosity may originated from the magnetosphere emission \citep{camp97,camp95}.

\section*{Acknowledgments}
We are grateful to the anonymous referee for useful suggestions. We also thank X.-D.
Li, Philipp Podsiadlowski, and H. Tong for helpful discussions. This work was partly supported by the National Science Foundation
of China (under grant number 11573016), Program for Innovative Research Team (in Science and Technology)
in University of Henan Province, and China Scholarship Council.

\bsp

\label{lastpage}

\end{document}